\documentclass[twocolumn,aps]{revtex4}
\input{epsfig.sty}
\newcommand{\beq}{\begin{eqnarray}}
\newcommand{\eeq}{\end{eqnarray}}
\begin{document}
\title{Pulsar Kicks With Modified URCA and Electrons in Landau Levels}
\author{Ernest M. Henley\\
Department of Physics, University of Washington, Seattle, WA 98195\\
Mikkel B. Johnson \\
Los Alamos National Laboratory, Los Alamos, NM 87545 \\
Leonard S. Kisslinger\\
Department of Physics, Carnegie Mellon University, Pittsburgh, PA 15213\\ }

\begin{abstract} We derive the energy asymmetry given the proto-neutron
star during the time when the neutrino sphere is near the surface of the
proto-neutron star, using the modified URCA process. The electrons produced
with the anti-neutrinos are in Landau levels due to the strong magnetic
field, and this leads to asymmetry in the neutrino momentum, and a pulsar
kick. The magnetic field must be strong enough for a large fraction of the
eletrons to be in the lowest Landau level, however, there is no direct
dependence of our pulsar velocity on the strength of the magnetic field.
Our main prediction is that the large pulsar kicks start at about 10 s
and last for about 10 s, with the corresponding neutrinos correlated in the
direction of the magnetic field. We predict a pulsar velocity of
1.03 $\times 10^{-4} (T/10^{10}K)^7$ km/s, which reaches 1000 km/s if
T $\simeq  9.96 \times 10^{10}$ K.

\end{abstract}
\maketitle
\noindent
PACS Indices:97.60.Bw,97.60.Gb,97.60.JD
\vspace{1mm}


\section{Introduction}

   The creation of neutron stars, often called pulsars, through neutrino
cooling starting less than a second after the collapse of a massive star
has long been of interest, with a number of processes contributing to
neutrino production\cite{bw,fsb}. The strong interaction treatment of
these processes was refined by Friman and Maxwell using perturbative one pion 
exchange and short-range interactions\cite{fm}. 

   In recent years it has been observed that many pulsars move with much
greater velocities than other stars in our galaxy. This is called the
pulsar kick. See Ref.\cite{hp97} for a review. Pulsars with velocities of
more than 1500 km/s have been observed. There has been a great deal of 
theoretical
effort in attempts to explain the pulsar kicks. For many years a number
of investigations of asymmetries in the hydrodynamics of core collapse
have been carried out, and in recent years there have been a number of 
reviews\cite{mbh,jskmp,wlh,jmk}. Although some simulations have found
possible pulsar velocities of 1000 km/s or more, there is no clear
proof that one can get the observed large pulsar velocities by the initial 
core collapse. There have been several calculations of possible asymmetry
in the neutrinos produced in strong magnetic fields using the
URCA process\cite{dor} and other processes\cite{hp,hl} during the 
first few seconds when the neutrinosphere has a radius of about 40 km. 
However. the opacities and short mean free paths of neutrinos in the 
neutrino atmosphere reduces the emission, and these processes cannot 
account for the large pulsar kicks\cite{lq,al}. There have also been 
calculations of pulsar kicks resulting from oscillation to sterile 
neutrinos\cite{ks97,fkmp03,bomz,bom}, which can escape from the neutrino
sphere during this early period. The recent MiniBooNE experiment\cite{Mini}
with previous LSND resiults are not consistent with a single sterile neutrino,
 but allow models with two or more sterile neutrinos. 

   It has long been recognized that during the later period, when the
neutrinosphere radius has been reduced to about 10 km, the radius of the
protoneutron star, that the modified URCA process dominates the cooling
of the protoneutron star\cite{bw,fm}. It is also known that 
protoneutron stars have very large magnetic fields. In the presence of
such fields the electrons produced in the modified URCA process will
be in Landau levels\cite{jl,mo}

   In the present work we derive the asymmetric neutrino emissivity
during the period when the neutrinosphere is just within the protoneutron
star surface, and show that due to the electrons being in Landau levels
pulsar velocities consistent with observations are obtained. In the
present work the contribution from polarization of the nucleons is not
included. Our work shows that for about a 10 s period starting at about 10 s,
when the modified URCA process dominates and the radius of the neutrinosphere
is a fraction of a km less than the neutron star radius, the main neutrino 
emission is asymmetric, and during this period the high velocity pulsar kicks
are generated. One of us presented a preliminary version of this work at 
the CosPA 2006 Symposium\cite{lsk07}. Also, the process of temperature 
equlibrium for the electrons in a strong magnetic field is being 
completed\cite{mbj07}

Our paper is organized as follows.  In Sect. II we give the basic quantities 
in terms of which we calculate the emissivity.
In Sect. III we explain the important differences between our 
formulation with the electrons in the lowest Landau level giving asymmetric 
momentum emissivity and the previous ones with no kick.
As explained in Sect. IIIB, we incorporate the 
Landau wave function in the lepton trace and give the results of the 
calculation of the traces, explaining that with the contributing electrons
all moving in the direction of the magnetic field the intregration over
the direction of the neutrino momentum is not present.  
In Sect. IIID we give our results for the
asymmetric  emissivity that gives the pulsar kick. In Sect. IIID1 we 
evaluate the essential ingredients of 
proto-neutron star structure that is needed for obtaining our final 
numerical result, given in Eq. (26) and Fig. 4.  In Sect. IV we 
present our conclusions. In the Appendix we give details of the calculation
of the matrix elements, the nucleon and lepton traces, and the angular 
integrals.

\section{Modified URCA Process in a Strong Magnetic 
Field: Landau Levels}

\begin{figure}[h]
%\begin{center}
\epsfig{file=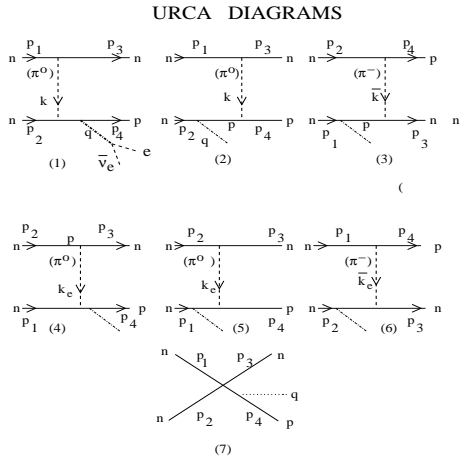,height=6cm,width=6.cm}
\caption{Modified URCA diagrams with OPE and a short-range n-n interaction}
{\label{Fig.1}}
%\end{center}
\end{figure}
The modified URCA process,
\beq
\label{1}
         n\;+\;n &\rightarrow& n\;+p\;+e^-\;+\bar{\nu}^e 
\eeq

\newpage
\noindent
for cooling of protoneutron stars has been treated in many publications
\cite{bw,fsb,fm}. In a detailed calculation\cite{fm} the one pion 
exchange (OPE) and a short-range interaction were used  for the nuclear 
interaction, illustrated in Fig 1.

%\vspace{3mm}
Diagrams (1,2,3) are for neutral and charged pion exchange, diagrams
(4,5,6) are the exchange diagrams, and 7 is the short-range diagram.
It can be shown that for our calculation of the asymmetric neutrino
emissivity the short-range n-n interaction is negligible, so we only consider
the OPE diagrams here.

   The OPE factor is given by the standard nonrelativistic form, while
the weak interaction used for the nucleons is the nonrelativiatic form of the 
standard model. As we shall show, only the weak axial interaction, $W_A$,
needs to be considered

\beq
\label{2}
         V_{OPE} &=& -\frac{f^2}{m_\pi^2}\tau^1 \cdot \tau^2 \vec{\sigma}_1
\cdot \vec{k} \frac{1}{k^2 + m_\pi^2}\vec{\sigma}_2\cdot \vec{k} \nonumber \\
   W_A &=& -\frac{G}{\sqrt{2}}g_A\chi_p^\dagger \vec{l}\cdot \vec{\sigma}
\chi_n \\
           l_\mu &=& \bar{\Psi}(q^e) \gamma_\mu(1-\gamma_5)\Psi(q^\nu) 
\nonumber \; ,
\eeq
with $\sigma,\tau$ the Pauli spin and isospin operators, (1,2) refer to the
two nucleons at pion vertices, 
$G=\frac{10^{-5}}{m_n^2},\;  g_A=1.26$, the $\chi$  are
the nucleon spinors, and the lepton wave functions are $\Psi(q^e),\Psi(q^\nu)$,
where  $q^e$ and $q^\nu$ are the electron and antineutrino momenta, 
respectively,  with the notation $q^e \equiv \vec{q}^e,
 q^\nu \equiv \vec{q}^\nu$. 
   Another point that should be mentioned is that the inverse modified
URCA processes which we neglect in the present work,
\beq
         n\;+\;p+\;e^- &\rightarrow& n\;+n\;+\nu^e \nonumber \\
            n\;+e^- &\rightarrow& n\;+\pi^-\;+\nu^e \nonumber \; ,
\eeq
will also have the electrons in Landau levels and thereby contribute to
the overall pulsar kick, as we show below.

\subsection{Landau Levels and Product Matrix Elements}

 The neutrino wave function is a Dirac spinor, while the electrons in a 
very strong magnetic field are in Landau levels, which
are similar to that of a Dirac particle in a plane wave state moving in the
z or -z direction, where $\hat{z}$ is the 
direction of the magnetic field, and the motion in the
transverse direction is of the form of a tightly bound state.  For a 
discussion of Landau wave functions see Refs\cite{jl,mo}. These states are
labeled by a principal quantum number, n, and spin and momentum.

  An essential aspect of the present work, is that a sizable fraction of the
electrons are in the lowest Landau level (n=0). One can estimate the 
fraction of electrons in this n=0 level from the energy gap between the n=0 
and n=1 levels, and T. The protoneutron stars that receive a
large velocity from the mechanism that we are considering have a large 
magnetic field strength, about $10^{15}$ to $10^{16}$ G at the star 
surface at a time of about 10 sec, when the neutrinosphere is near the 
surface of the protoneutron star. From expressions given in Ref\cite{mo},
the energy gap between the n=0 and n=1 levels is about 6.0 MeV. Our work,
shown below, concludes that a temperature T with kT about 8.59 MeV will yield
a pulsar velocity of about 1000 km/sec. From this we can use basic 
thermodynamics to estimate that the occupation probability of the n=1 state 
is about 0.5, and that the overall probability of the electron being in the 
n=0 level is about 40\%.

   A crucial point is that the electron in the n=0 level has its spin
in the -z direction, which we show below causes all of the emitted
neutrinos to be correlated in the z direction, while electrons in the
higher n levels have both helicities, and give no net pulsar kick.
Since we assume that all electrons produced via the modified URCA process
are in the n=0 Landau level, this introduces an error of a factor of 
aqpproximately 2, but as we shall see this is a trivial correction due 
to the very strong dependence of the resulting pulsar velocity on T, and
the uncertainty in the magnitude of T.

   The lowest Landau level, with n=0 has the form 
\beq
\label{3}
       \Psi(q^\nu) &=& u^{s,Dirac}(q^\nu) \equiv u^{s}(q^\nu) \nonumber \\
            \sum_s  u^{s}(q) \bar{ u^{s}}(q)&=& \not\!q +m \nonumber \\
       \Psi(q^e)&=& \psi^{Landau,n=0}(q^e_\perp,q^e_z,\phi) \\
 &=& i(\sqrt{\gamma})^{-1} e^{-(q^e_\perp)^2/(2 \gamma)} u^{-}(q^e)
\nonumber \; ,
\eeq
where $u^-$ is a negative helicity Dirac spinor, 
$\not\!q \equiv q^\mu \gamma_\mu$, and $\gamma = \delta m_e^2$; 
$\delta= B/(2B_c) \simeq B/8 \times 10^{13} Gauss\simeq $ 12.5 to 125 for
B=$10^{15}$ to $10^{16}$ Gauss.Using E+m $\simeq$ E, the wave functions have the conventional normalization 
$\int \bar{\Psi} \Psi =1$. The vector weak interaction is not included in 
our calculation as its contribution to the neutrino asymmetric emissivity is 
much smaller than the axial vector. The pulsar velocity is independent of 
the magnitude of B except for the n=0 occupation probability.

\newpage

   Although the z-component of the electron momentum can be in the +z or -z
direction, we shall show that there is no neutrino emission for the electron
moving in the -z direction, due to the vanishing of the lepton trace.

   For the calculation of the neutrino emissivity and the pulsar momentum
one needs the traces of the product matrix elements.  
The axial product matrix element, $|M_A|^2$, is obtained by taking the nucleon
traces over the product of the leptonic traces
times the square of the weak-strong product:
\beq
\label{4}
    |M_A|^2 &=& Tr(nucleon) | Tr(lepton)[l^{\dagger}_i l_j] \nonumber \\
  && [W_A(V_{OPE} + exchange)]_{ij}|^2
\; ,
\eeq 
with $V_{OPE}$ given in Eq(\ref{2}) and the (i,j) indices are the spatial
components of the lepton currents, defined in Eq.(\ref{2}). This is
treated in detail in the following section and in the Appendix.
\vspace{3mm}

   We introduce the following notation:
\beq
\label{5}
         k&=& p_1 -p_3\;,\;p=p_2-p_1\;,\;;k^e=k+p \nonumber \\
  \chi_i \chi_i^{\dagger} &\equiv& \Lambda_i \;= \;(1+\sigma \cdot P^{i})/2 \\
      A &=& (\frac{f}{m_\pi})^2\frac{G}{\sqrt{2}}g_A \frac{1}{\omega}\frac{1}
{(k^2+m_\pi^2)} \nonumber \\
    R(k)&=& \frac{k^2+m_\pi^2}{(k^{e})^2+m_\pi^2} \nonumber \; .
\eeq
The factor $R(k)$ is used in exchange matrix elements.

\section{ Neutrino Asymmetric 
Momentum Emissivity}

   With a strong magnetic field the electrons are in Landau levels. That is,
the motion of the electron transverse to the magnetic field is compressed,
and the electron is essentially a one-dimensional plane wave Dirac particle
moving in the direction of the magnetic field with energies given as Landau 
state energies. With a very strong magnetic field, such as near the surface
of the protoneutron star when the modified URCA process dominates neutrino
emission, the electron will fall into the lowest Landau level, with the
wave function given in Eq.(\ref{3}). The polarization of the nucleons is a 
very small effect\cite{bw}. The electrons being in Landau levels gives 
asymmetry to the neutrino emissivity. As we shall now discuss, the standard
formulation for obtaining the energy emissivity from the modified URCA
processes leads to the momentum of the emitted neutrinos correlated in the
B-field direction, and thus directly to the pulsar velocity. 

\subsection{Neutrino Asymmetric Emissivity}

   The neutrino emissivity is given in general form in many 
papers, e.g., see Ref\cite{fm}:
\beq
\label{6}
   e^\nu &=&\Pi^4_{i=1} \int \frac{d^3p^i}{(2\pi)^3}
\frac{d^3 q^\nu}{2 \omega^\nu(2 \pi)^3} 
 \int\frac{d^3q^e}{(2\pi)^3}
\nonumber \\
 && (2\pi)^4 \sum_{s_i,s^\nu} \frac{1} {2\omega^e_L}  \omega^\nu \mathcal{F}
M^{\dagger}_A M_A \\
&&  \delta(E_{final}-E_{initial}) \delta(\vec p_{final}-
\vec p_{initial})  \nonumber \\
    |M_A|^2 &=& M_{A-A}^{D-D} + M_{A-A}^{E-E} +M_{A-A}^{D-E} +M_{A-A}^{E-D} 
 \nonumber \; , 
\eeq
where $\mathcal{F}$ is the product of the initial and final Fermi-Dirac 
functions corresponding to the temperature and density of the medium.
We use the terminology that $M_A$ is the axial matrix element, which is 
given as the sum over the six Feynman-like diagrams shown in Fig. 1, but 
 keeping only the axial weak interaction, shown in Eq.(\ref{2});
$M^D_A$ stands for the direct diagrams 1,2, and 3 in Fig. 1, 
$M^E_A$ for diagrams 4,5, and 6,
$M_{A-A}^{D-D}$ is the product of $(M^D_A)^\dagger$ and $M^D_A$. 
$M_{A-A}^{E-E}$ and $M_{A-A}^{D-E},M_{A-A}^{E-D}, $ are analogously defined. 
The sum runs over the nucleon spins, $s_i$, and the neutrino spin. 
The electron is in the lowest Landau level, with 
its wave function  given in Eq(\ref{3}). 
\vspace{3mm}

   As in Refs \cite{bw,fsb,fm} the nucleons and the electrons are in
thermal equilibrium. In this present work we assume that the proton 
quickly reaches thermal equilibrium, while the process of the electron
state transforming to the n=0 Landau state does not interfere with the
proton reaching its Fermi momentum. Therefore we can use the values
for the magnitudes of the nucleon and lepton momenta derived 
in \cite{bw,fsb,fm}. 
\vspace{3mm}

 In the present paper we neglect the polarization of
the nucleons, so the $\Lambda_i =I/2$ (see Eq.(\ref{5})). Thus
the entire asymmetry of neutrino emission, which causes the pulsar kick
in the present work, arises from the electrons being in the lowest Landau 
level and the modified URCA process.
\vspace{3mm}

\subsection{Neutrino asymmetric momentum emissivity}

As seen from Eq.(\ref{6}) our calculation of the asymmetric neutrino
emissivity starts with the standard theory, except that the electron
is in the lowest Landau level, rather than in the usual Dirac state.
This turns out to be the crucial point. With the electron in the
lowest Landau level, the energy emissivity is also the projection
of the momentum emissivity along the $\hat{B}$ axis. That is, in the
following sections of the present paper we shall show that the 
emissivity given by Eq.(\ref{6}) is of the form
\beq
\label{7}
 e^\nu(q^\nu)^z &=& {\rm angular\; and\; energy\; integrals}\times
|M|^2 (q^\nu)^z \nonumber \\
  &=& P^{AS}(q^\nu)^z c\;=\; -p(ns) c \; ,
\eeq
where $Tr(lepton)Tr(nucleon)|M_A|^2 \equiv |M|^2 (q^\nu)^z$, c= speed of
light, $ P^{AS}(q^\nu)^z$ is the projection of the
momentum emissivity (momentum/volume/time) along the z=B direction, and
$p(ns)$ is the momentum/volume/time of the proto-neutron star. It
should be noted that the form equivalent to that of Eq(\ref{7}) has been
used by a number of authors studying pulsar kicks, such as Refs\cite{bomz,bom}.
\vspace{3mm}

 An essential ingredient in our framework is that electrons in the lowest
Landau level moving in the direction of the magnetic field are emitted 
while those moving in the opposite direction are not, which produces the
asymmetric momentum emissivity. This arises from the lepton traces
(see Eq.(\ref{4}) and Eqs.(\ref{A10},\ref{A11})) 
It is a straightforwrd exercise in Dirac algebra to show that for an 
electron in the lowest Landau level
$(I-\gamma_5) u^{-}(q^e) \bar{u}^{-}(q^e)]$ vanishes for 
$\hat{q}^e\cdot \hat{B}=\; -1$, so that we can write
\beq
 \int d^2 q^e_\perp Tr[l_i^{\dagger} l_j] & \simeq & 8 \pi E^e [(q^\nu)^j 
\delta_{i3} +(q^\nu)^i \delta_{j3} \nonumber \\
  &&-\delta_{ij} (q^\nu)^3]  (\hat{q}^e = \hat{B} = \hat{z}) \nonumber \; ,
\eeq
which is Eq.(\ref{A12}), derived in the Appendix. Note $(q^\nu)^i$ is the 
ith component of $q^\nu$, and $(q^\nu)^3=(q^\nu)^z$. This shows that when 
deriving the momentum asymmetry we can assume that the electrons in the 
lowest Landau levels are moving in the direction of the magnetic field.
We shall take $\vec{q}^e$ in the z (B) direction in the remainder of this 
paper.

   An important result of our work is that the pulsar velocity does not 
depend directly on the strength of the magnetic field, as is indicated in
Eq(\ref{A12}). This can be seen from the fact that the integral over the 
transverse components ofcthe electron in the lowest landau level, 
$\int dq^e_\perp q^3_\perp e^{- (q^e_\perp)^2/2\lambda} = 2\lambda$, 
gives a factor of $\lambda$
which cancels the factor of $1/\lambda$ from the square of the electron
wave function (see Eq.(\ref{3}). Therefore the only dependence of our
pulsar velocity on the magnetic field strength is the requirement that it 
is large enough that a sizable fraction of the electrons are in the n=0
Landau level.

\vspace{3mm}

 Therefore, with the electrons produced with the neutrinos in the modified
URCA process undergoing a transition to the lowest Landau level in the
very strong magnetic field near the protoneutron star, the standard
theory for energy emissivity also gives the emitted momentum of the neutrinos,
and therefore the recoil velocity of the resulting neutron star. However,
there is a great deal to discuss, including the calculation of the
various traces and the effective volume and time for the modified URCA
process to give a pulsar kick.
\vspace{3mm}
 
We start with the calculation of the traces of $|M_A|^2$.
As an example, let us calculate the direct axial matrix element.
Writing the product of matrix elements needed for the emissivity, as shown
in Eq.(\ref{6}), and using the notation given in Eq.(\ref{5}), with
$M_1$, $M_2$, and $M_3$ the 1, 2, and 3 diagram in Fig. 1,
\beq
\label{8}
    M1 + M2 &=& A\chi_3^\dagger \sigma \cdot k \chi_1 \chi_4^\dagger
[\sigma \cdot k,\sigma \cdot l]_+ \chi_2 \nonumber \\
        &&=2A l\cdot k\chi_3^\dagger \sigma \cdot k \chi_1
\chi_4^\dagger \chi_2 \nonumber \\
     M_3 &=& -2A \chi_3^\dagger \sigma \cdot k \sigma \cdot l \chi_1
\chi_4^\dagger \sigma \cdot k \chi_2 \\
    M_A^D &=& M1 +M2 +M3= 2A l\cdot k \chi_3^\dagger \sigma \cdot k \chi_1
  \nonumber \\
  && \chi_4^\dagger \chi_2 -\chi_3^\dagger\sigma \cdot k \sigma \cdot l\chi_1 
\chi_4^\dagger \sigma \cdot k \chi_2 \nonumber \; .
\eeq
Further results and the calculation of the $M_{A-A}^{D-D}$, $M_{A-A}^{D-D}$,
and $M_{A-A}^{D-E}$ traces of the product matrix elements are given in the 
Appendix. It is important to note that the Landau wave function is contained
in the lepton current, $l_i$. 
\vspace{3mm}

From the Appendix, Eqs.(\ref{A6},\ref{A9},\ref{A12}), and defining
$M^{DE}_{AA}\equiv M^{D-E}_{A-A}+M^{E-D}_{A-A}$
\beq    
\label{9}
  \int d^2 q^e_\perp M_{A-A}^{D-D} &=& \int d^2 q^e_\perp Tr[l_i^\dagger l_j]
A^2 (k^2 k_i k_j +k^4 \delta_{ij}) \nonumber \\
  &=& 8 \pi  A^2 E_e k^2 (-(q^\nu)^3 k^2+ 2(q^\nu)^i k_i k_z) \nonumber \\
    M_{A-A}^{E-E} &=& R(k)^2 M_{A-A}^{D-D}(k\rightarrow k^e) \\
   \int d^2 q^e_\perp M^{DE}_{AA} &=& -\int d^2 q^e_\perp Tr[l_i^{\dagger} l_j]
A^2 R(k) \nonumber \\ 
    && [-\frac{5}{2} k \cdot k^e (k_ik_j^e+k_i^e k_j) +2k_i^e k_j^ek^2 
\nonumber \\
 &&+2k_i k_j(k^e)^2 -(k \times k^e)_i (k \times k^e)_j \nonumber \\
   &&+(k \cdot k^e)^2 \delta_{ij} ] \nonumber \\
  &=&-8\pi A^2 R(k)E_e[-5 k \cdot k^e(k^e_z k \cdot q^\nu \nonumber \\
   &&+k_z k^e\cdot q^\nu) -2q \cdot k\times k^e (k\times k^e)_z 
\nonumber \\
  &&+4 k \cdot q^\nu k_z(k^e)^2+4 k^e \cdot q^\nu k^e_z (k)^2 \nonumber \\ 
     && +3(q^\nu)^z((k \cdot k^e)^2 - k^2(k^e)^2)] \nonumber \; .
\eeq

The next step in the calculation is to carry out the nucleon angular 
integrals.

\subsection{Angular integrals with k and k$^e$ angles independent}

 Following the prescriptions of Ref\cite{fm} the magnitudes of the
momenta are given by the Fermi momenta, so one only does integrals over
the energies  and angles of the momenta, and the momentum transfer of the
pion in the direct term, $\vec{k}=\vec{p}_1-\vec{p}_3$ is introduced as an
independent vector by inserting
\beq
       \int d^3 \delta(\vec{k}-\vec{p}_1+\vec{p}_3)=1 \nonumber \; .
\eeq

For the D-D (direct-direct) term, the only angular integral is over the
direction of k. For the E-E and D-E terms, however, one must deal with
$\vec{k}^e=\vec{k}+\vec{p}$ (see Eq.(\ref{5})). Using the $\delta$-functions,
one can see that neither $\vec{k}^e$ nor $\vec{p}$ are completely independent
of $\vec{k}$, but with the proton momentum being smaller than the neutron,  
following arguments in Ref\cite{fm}, it is a good approximation to assume 
that $\vec{k}^e$  and $\vec{k}$  are independent.
\vspace{3mm}

Using the results given in Eq.(\ref{9}) for the traces of the products
of the matrix elements, the integrals given in 
Eq.(\ref{A13}), and the approximations $k = k^e$ and $R(k)=1$, with
 the notation $\int\int$ = angular integrals over $\vec{k}$, $\vec{k}^e$, and
$q^e_\perp$  one finds
\beq
\label{10}
  \int\int M^{D-D}_{A-A} &=& \int\int M^{E-E}_{A-A}= - \frac{128}{3} 
\pi^2 A^2 k^4 (q^\nu)^z q^e \nonumber \\
     \int\int M^{DE}_{AA}&=& \frac{32}{9} \pi^2 A^2 k^4 (q^\nu)^z q^e
  \; .
\eeq

 Recognizing that the nucleons are in thermal equilibrium, the magnitudes 
of the nucleon momenta are given by the Fermi momentum, $p_F$, and therefore
the integrals over the magnitude of the four nucleon momenta are carried
out via delta functions. Dropping the $E^\nu$ terms, this gives the final 
result for the traces and integrals over the axial product matrix element

\beq
\label{11}
   \int\int |M_A|^2 &=& -0.81 \times 10^3 A^2 q^e (q^\nu)^z p_F^4 
 \nonumber \\
        &=&  -0.81 \times 10^3  (\frac{f}{m_\pi})^4 \frac{G^2 g_A^2}{2\omega^2}
\nonumber \\
   &&(\frac{p_F^2}{p_F^2+m_\pi^2})^2 q^e (q^\nu)^z \; .    
\eeq
\newpage

\subsection{Neutrino asymmetric emissivity}

 From Eq.(\ref{6}), based on the general ideas of 
Ref\cite{fm}, but with our formalism for the phase space integrals, 
with the angular integrations given by Eqs.(\ref{10},\ref{11}) in subsection 
III.C, we obtain the neutrino asymmetric emissivity: 
\beq
\label{12}
   (\epsilon^{AS})^\nu &=& 2 \int\int |M_A|^2 \frac{(m_n^*)^3 m_p^*}{(2 \pi)^9}
 p_F(e) I \; .
\eeq

The energy integrals, I, are the same as those in Ref\cite{fm}; and
with $q^e=85 MeV$, and $q^\nu = 4.7 kT$, we find:
\beq
\label{13}
    I &=& 9.04 \times 10^2 (kT)^8
\eeq
From Eqs.(\ref{11},\ref{12},\ref{13}) we obtain 
\beq
\label{14}
  \epsilon^{AS} &\simeq& 0.64 \times 10^{21} (\frac{T}{10^9 K})^7 
{\rm erg\; cm^{-3}\; s^{-1}}\nonumber \\
            &=& p_{ns} c ({\rm volume^{-1}\;time^{-1}}) \; ,
\eeq
where $p_{ns}$ is the momentum given to the neutron star, the volume
is the active region for the modified URCA process and the time represents
the time interval. $p_{ns}$ has the magnitude of $\epsilon^{AS}$/c in the
opposite direction of the net neutrino momentum.

   Due to the short mean free path of neutrinos within the neutrinosphere,
the main asymmetric emission from the process we have proposed will take
place in the volume between the neutrinosphere and the protoneutronstar
surface during the period when the neutrinosphere is just within the
protoneutronstar surface. At this time the temperature is expected to
be in the range $10^9K<T<10^{11}K$ for a period of 10s starting at about
10s. Taking the neutrinosphere and protoneutron star to have radii $R^\nu$ 
and $R_{ns}$, respectively, in km units, we find for the momentum 
given to the pulsar for this period of 10-20 s
\beq
\label{15}
   p_{ns} &\simeq& 0.43 \times 10^{27}(\frac{T}{10^9 K})^7 
\nonumber \\
        &&(R_{ns}^3-(R^\nu)^3) {\rm \;gm\;cm\;s^{-1}} \; ,
\eeq
\vspace{3mm}
where we use the effective volume for neutrino emission as $V_{ns}
-V_{\nu-sphere}=f\frac{4 \pi}{3}(R_{ns}^3-(R^\nu)^3)$., where f is the 
fraction of neutrinos which escape without striking the neutrinosphere for
various positions of emission. A rough quess is 0.5, and by integrating 
$q^\nu_z$, where z is defined in the direction of B, we find that f=0.52.

\newpage

\subsubsection{Radius of neutrinosphere during modified URCA emission}

   The final step in our derivation is to estimate the volume in which the
neutrino emission takes place with the modified URCA process in a strong
magntic field.  Referring to Eq.(\ref{23}), we must find the radius 
of the neutrinosphere, $R^\nu$, assuming that the radius of the protoneutron
star at this time is 10 km.

   To do this we use the Spherical Eddington model, which has been used by
a number of authors to study  the neutrino atmosphere associated with the
creation of a pulsar\cite{ss,jr,bomz,bom}. We follow the recent method
of Barkovich et. al.\cite{bom} to find the neutrinosphere radius, and
estimate the time and temperature during which our process is taking place.
See Ref\cite{plpsr}for a review of the evolution of the birth of a neutron
star.

  Our starting point is the energy-momentum tensor, $T^{\mu \nu}$ for the 
neutrinos with a distribution function $f^\nu(\vec{x},\vec{k},t)$ for each
type of neutrino, giving an energy density, $U$ and momentum density $\vec{F}$
\beq
\label{16}
    U&=& T^{00}=\int \frac{d^3k}{(2\pi)^3} k_0 f^\nu \nonumber \\
    F^i &=& T^{0i}= \int \frac{d^3k}{(2\pi)^3} k^i f^\nu \; .
\eeq

Making use of the Boltzmann equation, and recognizing that the neutrinos
have a very short mean free path, $\lambda^\nu$, the neutrino distribution
can be written in terms of the equilibrium distribution, 
$(f^{eq})^\nu(\vec{x},\vec{k})$ as
\beq
\label{17}
   f^\nu &\simeq& (f^{eq})^\nu -\lambda^\nu \hat{k}\cdot \Delta (f^{eq})^\nu 
\nonumber \\
    (f^{eq})^\nu &=& \frac{1}{1+e^{(k-\mu^\nu)/T}} \; ,
\eeq
which form we have used to obtain the integrals in Eq(\ref{16}).

The mean free path for each neutrino type in a medium of nucleons with
density $\rho$ can be written as
\beq
\label{18}
      \lambda^\nu &=& \frac{1}{\chi^\nu k^2 \rho} \; ,
\eeq
with the constant $\chi^\nu$ determined from the constants of the standard
weak interaction model and the cross sections for each neutrino type. Assuming 
the spherical Eddington model one finds for the energy and momentum densities,
including the electron, muon, and tau neutrinos, for equilibrium temperature T
\newpage

\beq
\label{19}
          U &=& \frac{7 \pi^2}{40} T^4 \\
         \vec{F} &=& -\frac{1}{36 \chi \rho} \frac{d T^2}{dr}\hat{r}
\nonumber \; .
\eeq

The time dependence of T can be found from energy-momentum conservation:
\beq
\label{20}
            \partial^\nu T^{0 \nu} &=& 0 \;{\rm or} \nonumber \\
            \partial_t U + \Delta \cdot \vec{F}& =& 0 \; .
\eeq

Using the equations of hydrostatic equilibrium, Barkovich et. al. showed that
the T can be obtained from the first order differential equation
\beq
\label{21}
  \frac{d\bar{T}}{dx} + b_c \frac{\bar{T}^2-a(x))}{\bar{T}^2 x^2 (1-a(x))}
&=& 0 \; ,
\eeq
with  $\bar{T}\equiv T/T_c$, $x=r/R_{ns}$ and  $a(x)$ depends on the 
constant $b_c$ and 
two other constants, as well as $\bar{T}$. We use the constants of 
Ref{\cite{bom}, with $T_c$ = 40 MeV, slightly altered for our expected 
luminoscity in the 10-30s interval when the modified URCA process dominates.

Our solutions are shown in figures 2 and 3.
\begin{figure}[ht]
%\begin{center}
\epsfig{file=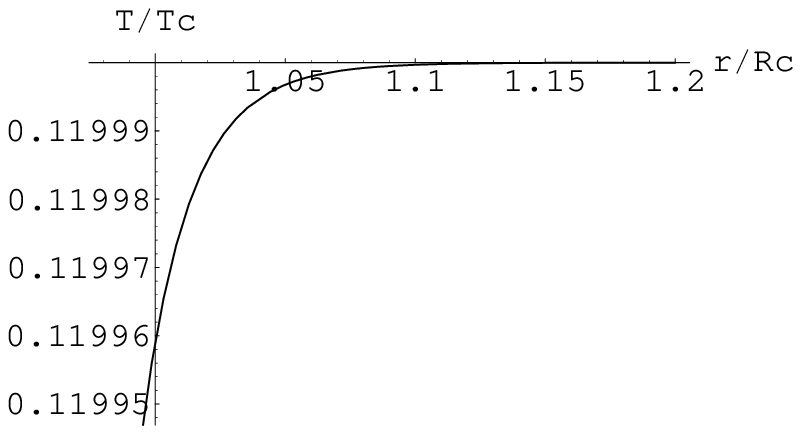,height=4cm,width=6.cm}
\caption{}
{\label{Fig.2}}
%\end{center}
\end{figure}

\begin{figure}[ht]
%\begin{center}
\epsfig{file=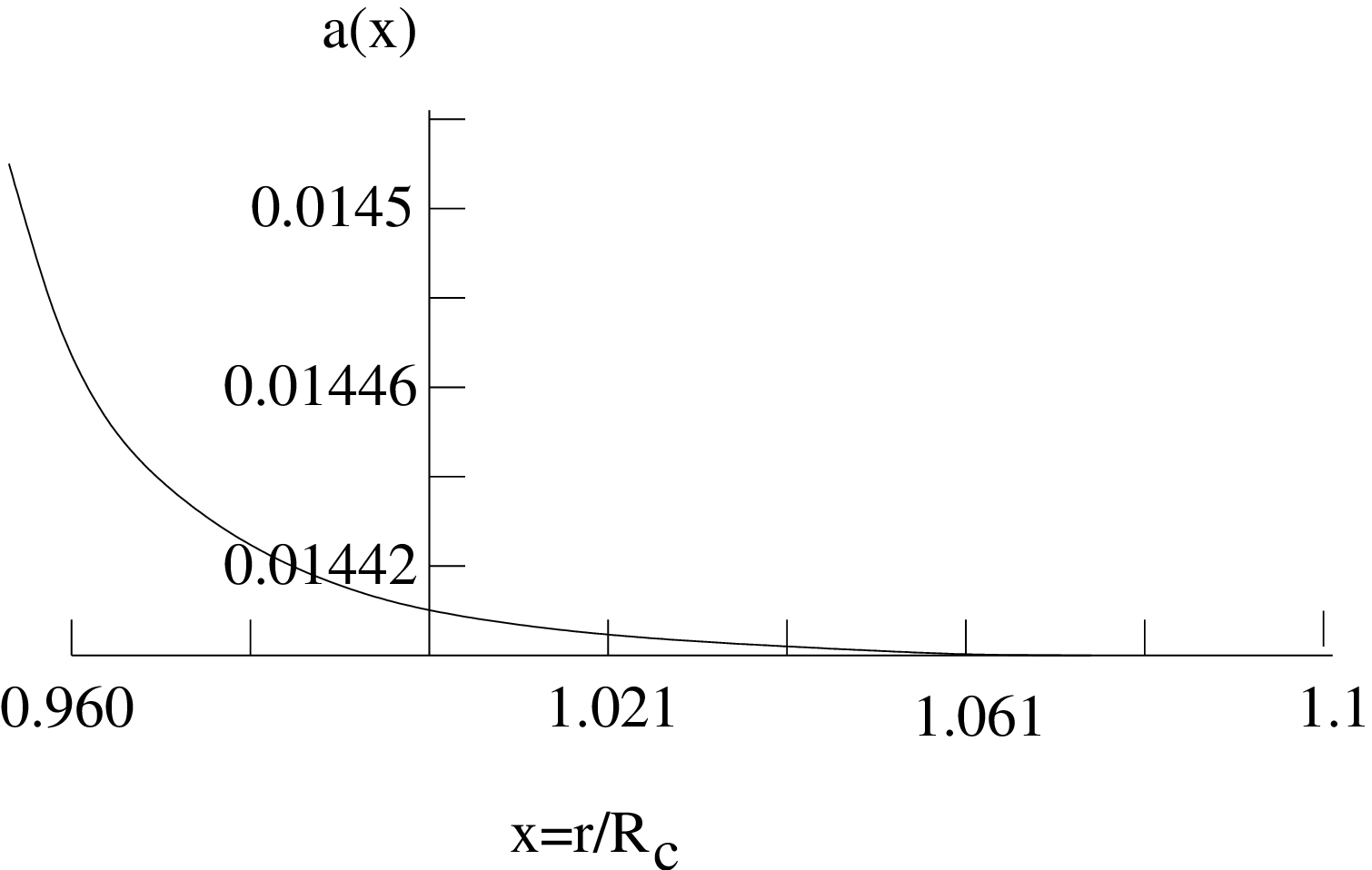,height=3cm,width=6.cm}
\caption{}
{\label{Fig.3}}
%\end{center}
\end{figure}

In our solutions we find that
$a(x) << 1, a(x)<< \bar{T}, a(x)\simeq \bar{T(x)}^2 (x \geq 1)$, and that the 
result for the mean free path is
\beq
\label{22}
         (\lambda^\nu)^{-1} &\simeq& \bar{T}(\bar{T}^2-a) cm^{-1} \; .
\eeq

With this solution the radius of the neutrinosphere can be obtained in terms
of $R_{ns}$ from the relation
\beq
\label{23}
 \int_{R^\nu/R_{ns}}^{\infty} \bar{T}(\bar{T}^2-a) dx &\simeq& 
\frac{2}{3}\frac{cm}{R_{ns}}\; .
\eeq
\vspace{3mm}

   Using our solutions we find that for $R_{ns}$ =10 km,
\beq
\label{24}
      R^\nu &\simeq& 9.96 {\rm \;km} 
\eeq
when the temperature is in the range $T\simeq 10^{10} K$, so that
from Eq(\ref{23}) the neutron star momentumm  is

\beq
\label{25}
    p_{ns} &\simeq& 5.14\times 10^{27} {\rm gm\;cm/s} (\frac{T}{10^9})^7
\nonumber \\
           &=& M_{ns} v_{ns} \; ,
\eeq
For a neutron star with the mass of the sun = $2 \times 10^{33}$ gm, including
a factor of 0.4 for the n=0 occupation probability,
\beq
\label{26}
    v_{ns} &=& 1.03 \times 10^{-4} (\frac{T}{10^{10}})^7 {\rm km\;s^{-1}} \; . 
\eeq
giving a velocity of v$\simeq$ 1000 km/s for T$\simeq 9.96\times 10^{10}$ K,
which is in the expected range.
Fig 4 illustrates the velocity of the pulsar as a function of T.

  Therefore we find that the modified URCA process can produce 
the observed velocities of 1000 km/s if T or more during this period if the
temperature is sufficiently high when the
neutrinosphere is slightly within the protoneutron star due to
the electrons being in Landau levels. These large pulsar kicks start about 10
s after the supernova collapse.

\begin{figure}[ht]
\begin{center}
\epsfig{file=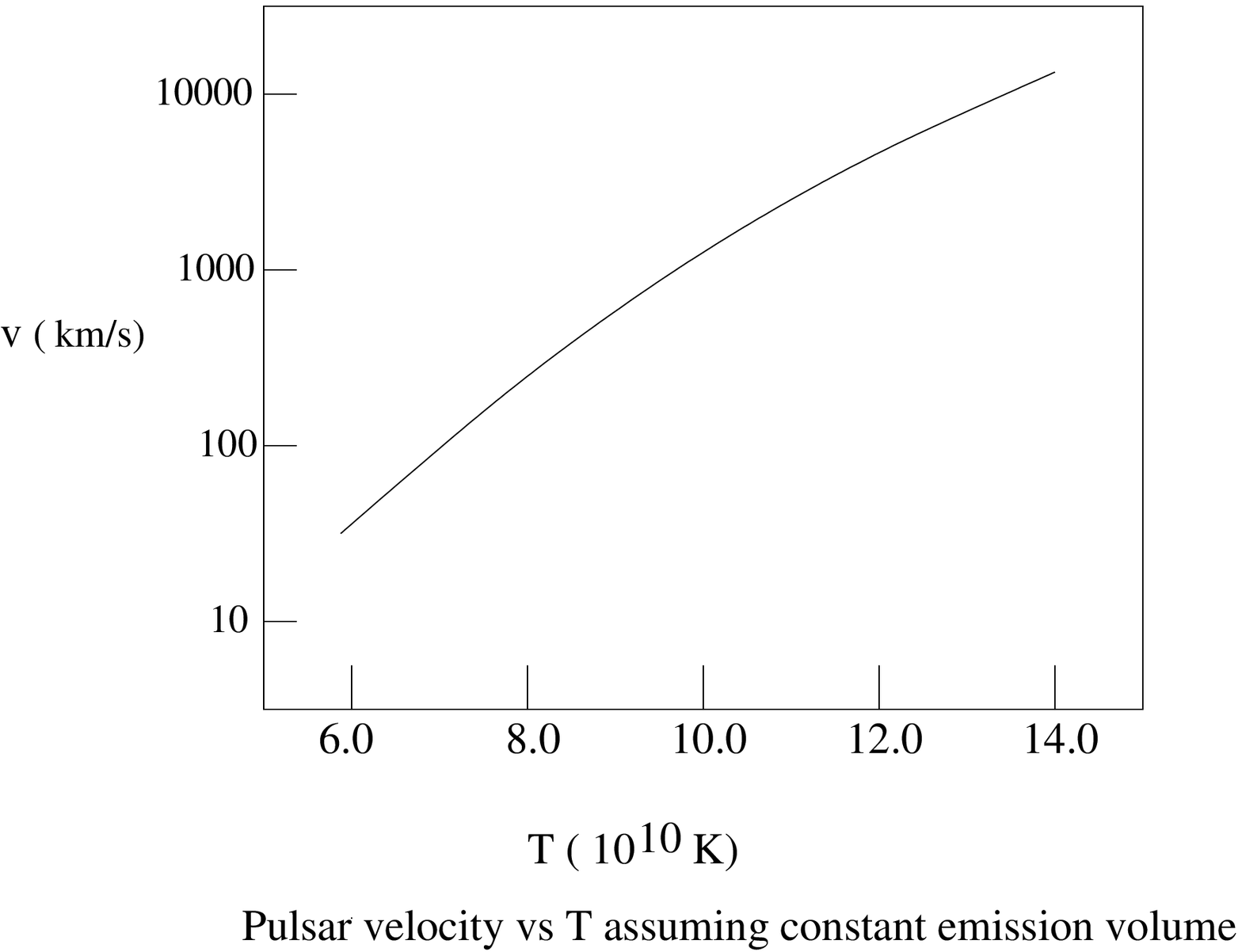,height=4cm,width=7cm}
\caption{}
{\label{Fig.4}}
\end{center}
\end{figure}
\newpage

\section{Conclusions}

 The basis of our present work is that for a magnetic field strength strong
enough for a sizable faction of electrons produced in the modified URCA
process to be in the lowest Landau lavel, all of the neutrinos emitted will 
be correlated with the direction of the magnetic field, giving a pulsar kick.
The resulting velocity depends strongly on the temperature, and only
indirectly on the magnetic field. Using expected values of $10^{15}$ to
$10^{16}$ Gauss for B, and the T for which we get a pulsar velocity of 
1000 km/s, we estimate that about 40\% of the electrons are in the n=0 
Landau level, which is satisfactory.

 The calculation of the asymmetric neutrino emissivity, and the resulting
pulsar velocity arising from electrons being in Landau orbits in the strong
magnetic field near the protoneutron star, is straightforward with the 
modified URCA process. Using the
standard properties of protoneutron stars in the 10-20 s time interval, we
find that with the electrons in Landau levels, the modified URCA process
can account for the measured pulsar kicks for $T > 9.96\times 10^{10}$K, 
with the neutrinosphere surface just inside the protoneutron star surface. 
Studies\cite{bom} give $T \simeq 10^{11} $K, or even greater  near the 
protoneutron star surface. We predict a strong correlation between the 
protoneutronstar T and pulsar velocity,
as well as a strong correlation between the direction of the
pulsar's velocity and the direction of B. Since the 
luminoscity of the pulsar is related to properties of the  protoneutron
star, such as strengths of B and T, one can also expect our result to
predict a correlation between v  and L of the pulsar
for high L. 

\begin{figure}[ht]
%\begin{center}
\epsfig{file=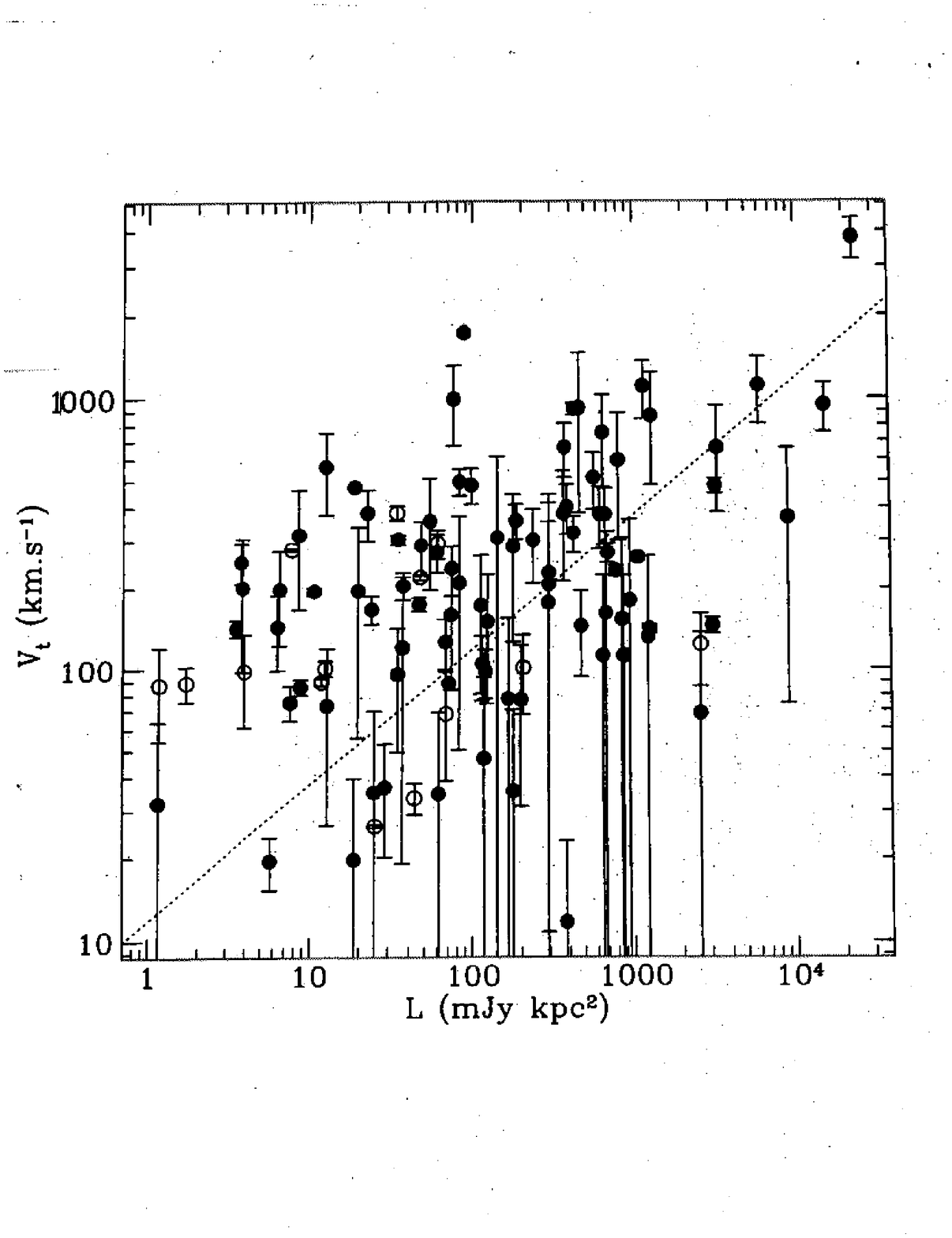,height=6 cm,width=8 cm}
\caption{Pulsar velocity vs luminoscity, Ref\cite{hp97}}
{\label{Fig.5}}
%\end{center}
\end{figure}

\newpage

In figure 5 the correlation between large L and large v that
has been observed is shown. It is a subject for future study for us to
see if our model is consistent with that observation.

Our unique prediction is that the main neutrino emission during this period 
when the neutrinosphere is just inside the the neutron star is almost entirely 
asymmetric, when the pulsar kicks should occur, with a strong
correlation with $E_\nu$. E.g., $E_\nu$ = 35 MeV$\leftrightarrow$ kT = 11 MeV
and v=2,000 km/s.

  It is interesting to consider the neutrinos observed in the 
Kamiokanda-II\cite{hir} and IBM\cite{bio} detectors from supernova SN1987A.
In the analysis of the neutrino data\cite{hir2}, in which the Kamiokanda-II
and IBM data are plotted as a function of time, there might be a gap in the
data, with additional neutrinos seen after 10 s (with the first 
neutrinos seen at about 1 s). See Fig. 6 for KAM II and IBM 
results\cite{hir2}.

 Although our agreement with the time gap seen in Fig 6 is not statistically 
significant, we predict asymmetric neutrinos start to appear at
about 10 s, correlated in the B direction. This should be
observable in future measurements of neutrinos from supernovae with
today's improved detectors, with neutrino energies determining the
pulsar velocities.

\begin{figure}[ht]
%\begin{center}
\epsfig{file=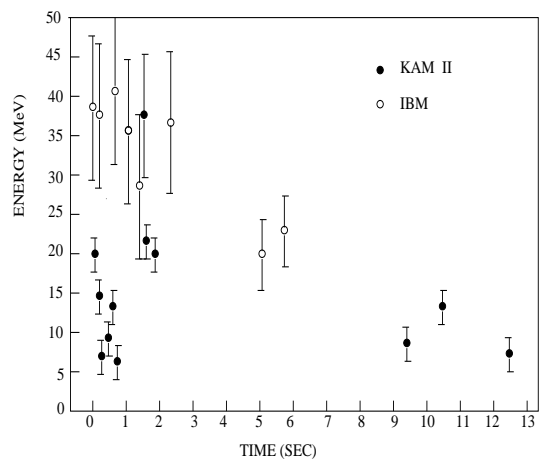,height=6 cm,width=7 cm}
\caption{Energies of neutrinos from SN1987A}
{\label{Fig.6}}
%\end{center}
\end{figure}

\section{Appendix}
\subsection{ Axial-Axial (A-A) Matrix elements}
  The axial OPE matrix element has a direct and an exchange part,
\beq
\label{A1}
      M_A &=& M_A^D + M_A^E \; ,
\eeq
corresponding the diagrams 1,2, and 3 for the direct and 4,5, and 6 for
the exchange in Fig. 1. Using the notation $\chi_n$ for the spinor of the
nucleon n (n=1,2 for the initial neutrons, n=3 for the final neutron, and 
n=4 for the produced proton), from Eq.(\ref{2},\ref{6}) 
\beq
\label{A2}
      M_A^D &=& 2 A [\vec{l}\cdot \vec{k} \chi_3^{\dagger}\sigma\cdot \vec{k}
\chi_1 \chi_4^{\dagger} \chi_2-\chi_3^{\dagger}\sigma \cdot\vec{k} 
\sigma \cdot l\chi_1 \nonumber \\
             && \chi_4^{\dagger} \sigma\cdot \vec{k} \chi_2] 
\eeq
\beq          
\label{A3}
        M_A^E &=&- R(k)  M_A^D (k \rightarrow k^e, 1\leftrightarrow 2)
\eeq

  Using $\chi_i \chi_i^{\dagger} \equiv \Lambda_i$, the D-D product matrix
 element is
\beq
\label{A4}
      M_{A-A}^{D-D} &=& 4 A^2 Tr[ l\cdot k l^{\dagger} \cdot k
(A1)(A2) +(A3)(A4)\nonumber \\
       && -l^{\dagger} \cdot k (A5)(A6) -l\cdot k (A7)(A8)] \; ,
\eeq  
 where, assuming that $\bar{k}=k-q \simeq k$, $\Lambda_n = I/2$ and with 
the notation $\vec{k}\rightarrow k$

\beq
\label{A5}
         (A1) &=& Tr[\sigma \cdot k \Lambda_3 \sigma \cdot k
 \Lambda_1] \; = \; \frac{1}{2} k^2 \nonumber \\
          (A2) &=& Tr[\Lambda_2 \Lambda_4] \; = \; \frac{1}{2} \nonumber \\
          (A3) &=& Tr[\sigma \cdot l^\dagger \sigma \cdot k\Lambda_3
 \sigma \cdot k\sigma \cdot l \Lambda_1] \nonumber \\
             && = \; \frac{1}{2} l^\dagger_i  l_j k^2 \delta_{ij} 
 \nonumber \\
          (A4) & \simeq & (1A)\; =\; \frac{1}{2} k^2 \\        
          (A6) &=& Tr[\sigma \cdot k \Lambda_2 \Lambda_4]\; = 0
\nonumber \\
          (A8) &=& Tr[\sigma \cdot k \Lambda_4 \Lambda_2]\; = 0 \nonumber \; .
\eeq
From this we find
\beq
\label{A6}
   M_{A-A}^{D-D} &=&  A^2 Tr[l_i^{\dagger} l_j]k^2 (k_i k_j + k^2 \delta_{ij})
\\
   M_{A-A}^{E-E} &=&  A^2 R(k)^2 Tr[l_i^{\dagger} l_j](k^e)^2 (k^e_i k^e_j + 
 (k^e)^2 \delta_{ij}) \nonumber
\eeq

Using $\bar{k} \simeq k, \bar{k}^e \simeq k^e$,
the two direct-exchange product matrix elements can be written as (note
$ M_{A-A}^{E-D}= M_{A-A}^{D-E}(k\leftrightarrow k^e)$)
\beq
\label{A7}
      M_{A-A}^{D-E} &=& -4 A^2 R(k)\sum_{spins}[l^\dagger\cdot k
l\cdot k^e DE1  \\
   && + DE2 -l^\dagger\cdot k DE3 -l\cdot k^e DE4] \; ,\nonumber
\eeq
with (assuming $\Lambda_n = I/2$ for all nucleons)
\beq
\label{A8}
   DE1&=& \frac{1}{16}Tr[\sigma \cdot k \sigma\cdot k^e ] \; = \; 
\frac{1}{8} k \cdot k^e  \nonumber \\
   DE2&=& \frac{1}{16} Tr[ \sigma \cdot l^\dagger\sigma \cdot k 
\sigma \cdot k^e \sigma \cdot l \sigma \cdot k \sigma \cdot k^e]
\nonumber \\
    &=&\frac{1}{8}l^\dagger_i l_j[-k\cdot k^e(k_i i^e_j +k^e_i i_j)
+k_ik_j(k^e)^2 \nonumber \\ 
  && +k^e_ik^e_j(k)^2  -(k\times k^e)_i (k\times k^e)_j 
+\delta_ {ij}(k\times k^e)^2] 
\nonumber \\
   DE3 &=& \frac{1}{16}Tr[ \sigma \cdot k \sigma \cdot k^e
 \sigma \cdot l \sigma \cdot k^e]  \nonumber \\
    &=& \frac{1}{8}(2 k \cdot k^e l \cdot k^e-l\cdot k (k^e)^2 )\nonumber \\
   DE4 &=&\frac{1}{16}Tr[\sigma \cdot l^\dagger \sigma \cdot k  
  \sigma \cdot k^e \sigma \cdot k]  \nonumber \\ 
      &=& \frac{1}{8}(2 k \cdot k^e l^\dagger-l^\dagger\cdot k^e (k)^2 )
  \; .
\eeq

To obtain Eq.(\ref{A8}) we use the trace relationships
\beq
\label{tr1}
    Tr[\sigma \cdot A \sigma \cdot B \sigma \cdot C \sigma \cdot D]
&=& 2(A \cdot B C \cdot D  \\
   &&-A\times B \cdot C\times D) \; , \nonumber
\eeq

\beq
\label{tr2}
  && Tr[\sigma \cdot A \sigma \cdot B \sigma \cdot C \sigma \cdot D
 \sigma \cdot E \sigma \cdot F] \nonumber \\
  && = 2 E \cdot F (A \cdot B C \cdot D  -A\times B \cdot C\times D)\\
    && -2[A \cdot B E \times F \cdot C \times D + C \cdot D E \times F 
\cdot A \times B 
 \nonumber \\
   &&- ((A \times B)\times (C \times D))\cdot E \times F] \; .\nonumber
\eeq
From this, 
defining $M^{DE}_{AA} = M^{D-E}_{A-A}+ M^{E-D}_{A-A}$ we obtain
\beq
\label{A9}
  M^{DE}_{AA} &=& -A^2 R(k) Tr[l^\dagger_i l_j] [-\frac{5}{2} k\cdot k^e
(k_ik^e_j+k^e_ik_j) \nonumber \\
  && +2 k_ik_j(k^e)^2 +2 k^e_ik^e_j(k)^2 +(k\cdot k^e)^2 \delta_{ij}
\nonumber \\
  && -(k\times k^e)_i (k\times k^e)_j]\; .
\eeq

\subsection{Lepton Traces}

   The lepton trace, $Tr[l_i^{\dagger} l_j]$, from Eq.(\ref{2}), with the
neutrino in a standard Dirac state, $ u^{s}(q^\nu)$, and the
electron in the lowest Landau level, $\Psi(q^e)=i(\sqrt{\gamma})^{-1}) 
e^{-(q^e_\perp)^2/(2 \gamma)} u^{-}(q^e)$, with the electron momentum along 
direction of the magnetic field, its spin in the opposite direction,
 is defined as
\beq
\label{A10}
    Tr[l_i^{\dagger} l_j] &=& \frac{2}{\gamma}  e^{-(q^e_\perp)^2/ \gamma}
Tr[\gamma_i \not\!q^\nu \gamma_j (I-\gamma_5) \nonumber \\
     &&  u^{-}(q^e) \bar{u}^{-}(q^e)] \; .
\eeq
Using $m_e \ll E_e$, and dropping terms in $(q^0)^\nu$, which do not give 
rise to momentum asymmetry,
\beq
\label{A11}
   \gamma_i \not\!q^\nu \gamma_j|_{i \ne j} &=& -(q^i)^\nu\gamma_j
-(q^j)^\nu\gamma_i  \nonumber \\
   \gamma_i \not\!q^\nu \gamma_i &= & \not\!q^\nu-2 (q^i)^\nu \gamma_i \\
(I-\gamma_5) u^{-}(q^e) \bar{u}^{-}(q^e) &=& \frac{E^e}{2}[\gamma^o
      +\gamma^o \gamma^5 + \gamma^5\gamma^3-\gamma^3] \nonumber \\
    && (\hat{q}^e = \hat{B} =\hat{z}) \nonumber \\
  \int d^2 q_\perp \frac{1}{\gamma}  e^{-(q_\perp)^2/ \gamma} &=&2 \pi 
\nonumber \; .
\eeq

An essential part of our work is  that $ (I-\gamma_5) u^{-}(q^e) 
\bar{u}^{-}(q^e)$ vanishes if the electron momentum is opposite to the B-field,
and that only electrons in the lowest Landau level in the direction of the
B-field contribute to the emissivity.
 From Eqs.(\ref{A10},\ref{A11}) we find for the electron in the lowest Landau
level
\beq
\label{A12}
  \int d^2 q^e_\perp Tr[l_i^{\dagger} l_j] &\simeq& 8 \pi E^e
 [(q^\nu)^j \delta_{i3} +(q^\nu)^i \delta_{j3} \nonumber \\
    && -\delta_{ij} (q^\nu)^3] (\hat{q}^e = \hat{B} =\hat{z}) \; .
\eeq

\subsection{Angular Integration}

   Using $k=p_1-p_3, k^e = p_2-p_1$ as independent variables (see text
for discussion), and defining $ \int\int \equiv \int d\Omega_k\int 
\frac{d\Omega_{k^e}}{4\pi}$
 We need the following angular integrals
\beq
\label{A13}
        \int\int k_i &=& \int\int k^e_i \; = 0 \\
   \int\int k \cdot A k \cdot B &=& \frac{4 \pi k^2}{3} A \cdot B \nonumber \\
  \int\int k \cdot A k \cdot B k \cdot C k \cdot D &=& \frac{4 \pi k^4}{15} 
 (A \cdot B C \cdot D + \nonumber \\
          A \cdot C B \cdot D +A \cdot D B \cdot C) \nonumber \\
   \int\int k^2 &=& 4 \pi k^2 \nonumber \\
    \int\int (k_z)^2 &=& \frac{4 \pi}{3} k^2 \nonumber \\
  \int\int (k \cdot k^e)^2 &=& \frac{4 \pi}{3}  k^2 (k^e)^2 \nonumber \\
  \int\int k \cdot k^e k_z k^e_z &=& \frac{4 \pi}{9}  k^2 (k^e)^2 \nonumber \\
  \int\int ((k \times k^e)_z)^2 &=& \frac{8 \pi}{9}  k^2 (k^e)^2 \nonumber \\
  \int\int k \cdot k^e k_z k^e \cdot q &=& \frac{4 \pi}{9}  k^2 (k^e)^2 q_z
\nonumber \\
  \int\int  k^e_z k^e \cdot q &=& \frac{4 \pi}{3} (k^e)^2 q_z \nonumber \\
 \int\int k \times k^e \cdot q ( k \times k^e)_z &=& \frac{8 \pi}{9}  
k^2 (k^e)^2 q_z \nonumber \\
  \int\int (k \times k^e)_i k \cdot k^e &=& 0 \nonumber \; .
\eeq

\Large{{\bf Acknowledgements}}\\
\normalsize

This work was supported in part by DOE contracts W-7405-ENG-36 and 
DE-FG02-97ER41014. The authors thank Sanjay Reddy and Richard Schirato
for helpful discussions. EMH and LSK thank LANL for hospitality while
part of this work was carried out.

%\begin{references}


\begin{thebibliography}{99}
\bibitem{bw}J.N. Bahcall and R.A.Wolf, Phys. Rev. {\bf 140}, B1452 (1965)
\bibitem{fsb}E.G. Flowers, P.G. Sutherland and J.R. Bond, Phys. Rev. {\bf D12},
315 (1975)
\bibitem{fm}B.L. Friman and O.V. Maxwell, ApJ {\bf 232}, 541 (1979)
\bibitem{hp97}B.M.S. Hansen and E.S. Phinney, astro-ph/9708071, Mon. Not. R.
Astron. Soc. {\bf 291}, 569 (1997).  
\bibitem{mbh}J. Murphy, A. Burrows and A. Heger, ApJ {\bf 615}, 460 (2004)
\bibitem{jskmp}H-Th. Janka, L. Sheck, K. Kifonidis, E. M$\ddot{u}$ller and
T. Plewa, astro-ph/0408439
\bibitem{wlh}Chen Wang, Dong Lai and J.L. Han, Astrophysis.J {\bf 656}, 399
(2007)
\bibitem{jmk}H-Th. Janka, A. Marek and F-S. Kitaura, astro-ph/0706.3056
\bibitem{dor}O.F. Dorofeev et. al., Sov. Astron. Lett. {\bf 11}, 123 (1985)
\bibitem{hp}C.J. Horowitz and J. Piekarwitz, Nucl Phys. {\bf A640}, 281 (1998)
\bibitem{hl}C.J. Horowitz and G. Li, Phys. Rev. Lett. {\bf 80}, 3694 (1998)
\bibitem{lq}D. Lai and Y-Z Qian, astro-ph/9802345
\bibitem{al}P. Arras and D. Lai, Phys. Rev. {\bf D60}, 043001 (1999)
\bibitem{ks97}A. Kusenko and G. Segre, Phys. Lett. {\bf B 396}, 197 (1997)
\bibitem{fkmp03}G. M. Fuller, A. Kusenko, I. Mocioiu and S. Pascoli,
Phys. Rev. {\bf D 68}, 103002 (2003)
\bibitem{bomz}M.Barkovich, J.C. D'Olivo, R. Montemayor and J.F. Zanella,
Phys. Rev. {\bf D 66}, 123005 (2002)
\bibitem{bom}M.Barkovich, J.C. D'Olivo and R. Montemayor, hep-ph/0503113
\bibitem{Mini}A.A. Aguilar-Arevalo et al (MiniBooNE Collaboration),
Phys. Rev. Lett. {\bf 98}, 231801 (2007)
\bibitem{jl}M.H. Johnson and B.A. Lippman, Phys. Rev. {\bf 76}, 828 (1949)
\bibitem{mo}J.J. Matese and R.F. O'Connel,  Phys. Rev. {\bf 180}, 1289 (1969)
\bibitem{lsk07}L.S. Kisslinger, hep-ph/0612221 to be published in Mod. Phys.
 Lett. A (2007)
\bibitem{mbj07} A study by E.M. Henley, M.B. Johnson and L.S. Kisslinger
of the equilibrium conditions for the modified URCA process in a 
protoneutron star is in preparation.
\bibitem{ss}P.J. Schinder and S.L. Shapiro, ApJ {\bf 259}, 311 (1982)
\bibitem{jr}H-T. Janka and G.G. Raffelt, Phys. Rev. {\bf D 59},023005 (1998)
\bibitem{plpsr}M. Prakash, J.M. Lattimer, J.A. Pons, A.W. Steiner and
S. Reddy, astro-ph/0012136, Lect.Notes Phys. {\bf 578}, 364 (2001)
\bibitem{hir}K. Hirata et.al., Phys. Rev. Lett. {\bf 58}, 1490 (1987)
\bibitem{bio}R.M. Bionta et. al., Phys. Rev. Lett. {\bf 58}, 1494 (1987)
\bibitem{hir2}K. Hirata et.al., Phys. Rev. {\bf D38}, 448 (1988)


%\end{references}
\end{thebibliography}
\end{document}